# Dr. Harold Whichello: medicine and astronomy in Cheshire

**Jeremy Shears and Theresa Hull**

**Abstract**

Dr. Harold Whichello (1870-1945) was a Cheshire General Practitioner and an enthusiastic amateur astronomer. He joined the British Astronomical Association in 1898 and undertook observations for its Lunar, Solar and Variable Star Sections using a 6-inch Wray refractor. He also contributed lunar occultation predictions and comet ephemerides to its Computing Section.

**Introduction**

In 1907 Paul Henri Stroobant (1868-1936) of the Royal Observatory of Belgium (ROB) (1) published what was effectively the first global astronomical directory, "*Les Observatoires Astronomiques et les Astronomes*" (2) (Figure 1a). This listed some 260 astronomical observatories, institutes and societies, together with the names of about 1,500 astronomers. The names were gleaned by ROB staff members from astronomical journals as well as publications of observatories and societies. In an attempt to identify the most active amateur observers, Stroobant and his colleagues also contacted officers of astronomical societies, including the Liverpool and Manchester Astronomical Societies, as well as BAA Section Directors (3). For each observatory, the place name and principal observers or researchers were given, along with a description of the instruments used and chief areas of research, supplemented in some cases with a few historical notes. As one might expect, the list was graced by the names of world famous research institutions like the Royal Observatories at Greenwich and Edinburgh, the Meudon Observatory in France and the US Observatories at Harvard, Lick and Yerkes. A number of private individuals were also listed, including Camille Flammarion (1842-1925) of Juvisy in France, W.F. Denning (1848-1931) of Bristol and Isaac Roberts (1829-1904) of Crowborough, amongst others. However, alongside these famous names, there appears the following entry (Figure 1b):

> **Chester** (*Angleterre*)
>
> **Whichello, H., Dr.   The Mount, Tatterhall, Chester**
>
> *Lune.*
>
> *Réfracteur de 152 millimètres d'ouverture*

Tattenhall (misspelt in the book as "Tatterhall") is a village in the county of Cheshire, some 14 km south-east of Chester and is home to one of the present authors (TH) and a short distance from the other (JS). We were naturally intrigued by the mention of a local amateur astronomer who, for the period, possessed a fairly large telescope. So who was Dr. Whichello (Figure 2), what was his contribution to astronomy and what did he observe with his 6-inch refractor? This paper explores these questions in an attempt to shed some light on a little known amateur astronomer.

**Early life and education**

Harold Whichello was born on 31 March 1870 at Linslade, Buckinghamshire, which is now in the county of Bedfordshire. His father, Henry Whichello (1829-1908), was a Timber Merchant. His mother, Amy E. Whichello (1843-1915), was originally from Wollaston in Northamptonshire (4). Harold's brother, Norman, was two years his junior (5).

Harold was educated at Bedford Modern School (Figure 3) between 1882 and 1886 (6). On leaving school, and wishing to improve his language skills, he travelled firstly to the University of Gőttingen in Germany where he undertook a course in astronomy and then to the University of Geneva. His grasp of German and French was to be great benefit in later life when he read widely in both languages. On returning to England he studied medicine St. Thomas's Hospital in London, becoming LRCP and MRCS.

**A Cheshire GP**

Having qualified as a doctor in 1894 Whichello briefly served as a ship's surgeon with the Shaw, Savill & Albion Company's *SS Coptic*, a steamer which was permanently engaged in the New Zealand service via Australia. Evidently the long voyages provided ample opportunity to conduct some basic medical research for Whichello later wrote a letter to the *Lancet* (7) describing the case of a female passenger who presented with symptoms of diabetes. The standard Fehling's test for sugar in the urine yielded a positive result, which would have been indicative of diabetes. However, realising that the lady was also receiving two medicines in connection with tuberculosis, namely sulphonal and creosote (8), he wondered whether these might be interfering with the Fehling's test. Sure enough, withdrawing both sulphonal and creosote resulted in a negative Fehling's test. Reintroducing each separately elicited a positive reaction with only one: the sedative sulphonal. Considering that sulphonal was widely used in medicine at the time, Whichello naturally wanted to alert other doctors to his findings.

Upon returning to England, Whichello set up in General Practice in the village of Tattenhall where he resided at The Mount, an imposing property on the High Street (Figure 4) (9). In 1902 he married Blanche Mary Read (1872-1934), daughter of Colonel Alfred Read, a wealthy ship-owner from Chester (10), and the couple had two daughters: Marie Blanche (11) and Eva Florence (12). Harold Whichello's parents also moved to Tattenhall to be near their son (13).

In his early medical career, the young doctor, not yet 30 himself, was involved in one of the most tragic events ever to have happened in Tattenhall. In the late afternoon of Monday 15 May 1899 two brothers from the village, Thomas and George Cooke, aged 28 and 22 respectively, were returning through the Cheshire lanes from the agricultural market at Beeston Castle in an open horse and trap, a journey of approximately 9 km. Just over halfway home, the heavens opened and they encountered a 'terrific thunderstorm, accompanied by vivid flashes of lightning and a downfall of phenomenally large hailstones'. The brothers were struck by lightning and fell simultaneously out of the back of the cart. A short time later they were found by a local farmer lying in the road with the bench seat of the cart still in place under them. Whichello was summoned from the village arriving about 15 minutes later, but after an unsuccessful attempt at resuscitation by artificial respiration, he pronounced them dead. On further inspection he observed that: (14)

"There was no mark at all upon the elder brother, and it is thought that George [the younger brother], who was driving, was struck by the lightning, and that probably the fatal current passed through his body into his brother. George's face and breast were severely singed, the lightning having burned holes in his waistcoat, shirt, and singlet, and his collar stud was driven into his neck".

It must have been a harrowing sight. Moreover as Whichello remarked "[a]n odour of burnt flesh pervaded the whole width of the road" (15). The horse ran home to Tattenhall Hall, where the young men's father who, "uneasy at the strange return of the horse and cart, had walked down the village to inquire into the matter, and the sad news was broken to him as gently as possible" (14).

An inquest was held at the Bear Hotel in Tattenhall and the brothers were subsequently buried in the churchyard of St. Alban's Church. The High Street was lined with mourners for the occasion as the whole village turned out to express their condolences.

An equally tragic medical event recorded by Dr. Whichello involved one of his friends in Geneva who he reported "was slowly drugged to death by a chemist's assistant putting ten times as much arsenic into a tonic as the prescriber intended" (16). The explanation was that "though the assistant was accustomed to the metric system from boyhood, he made the simple mistake of displacing the decimal point one place". Whichello was clearly not at all enamoured with the metric system. He believed that it was far too "easy to mistake 0.002 g for 0.02 g, especially with the calligraphy chemists usually have to contend with". Moreover he maintained that it was harder for medics to remember decimal doses when prescribing drugs: "vulgar fractions present a definite picture to [the] mind, but not so decimals". Thus a decimal dose such as "0.015 g to 0.06 g is as difficult to call to mind as an address in New York".

Although the Whichello family appear to have been very well integrated into Tattenhall village life, taking part in a variety of social events and clubs, in 1914 they moved to Dr. Whichello's new practice at Heswall on the Wirral (17) and in 1923 they again moved to Lache Lane in Chester.

**Observational astronomy**

Of critical importance in the life of the young Harold Whichello was his reading of Agnes Clerke's (1842-1907) *A Popular History of Astronomy during the Nineteenth Century* (18). Thus captivated, he later pursued a course in astronomy at Gőttingen University before proceeding to medical school in London. For a period thereafter his hobby appears to have taken a back seat relative to his medical studies and early career progression. In 1898, however, he was elected a member of the BAA (19) being proposed by another amateur astronomer from Cheshire, F.W. Longbottom (1850-1933). Longbottom was an original member of the BAA and was well-known for his infectious enthusiasm for astronomy (20). He operated several telescopes from his garden in Chester and was also listed in Stroobant's "*Les Observatoires Astronomiques et les Astronomes*", Longbottom's name immediately preceding the Whichello entry (Figure 1b). Longbottom was a keen astronomical photographer and went on to serve as Director of the BAA Photographic Section 1906-1926. The two became firm friends and kept in contact for many years. Whichello's seconder on his BAA application form was the Association's Secretary, William Schooling (1860-1936) (21).

Whichello's first intervention in BAA affairs was a letter he submitted to the *Journal* within just a few months of joining on "An Instrument Loan Fund" (22). During the early years of the Association's existence there was a keen desire to establish an observatory in Regent's Park in London for the benefit of members and an Observatory Fund had been set up to provide appropriate finance for it (23). However, by 1898 the fund had not even reached one-half of its target and there were suggestions that the idea should be dropped. Whichello was of the opinion that if an observer was to be involved in a sustained programme of work, then his instrument really should be located at his home for ease of access. In that way even brief spells of good weather could be fully utilised. Many members objected to the inconvenience of the central London location of the proposed observatory. Whichello had similar concerns: (24) "The proposed observatory in Regent's Park may be of use to a few men living close to it, but to 99 per cent of the 1,100 members the place will be almost useless. The Liverpool society's observatory is very little used, though it contains a first-class 5-in. Cooke refractor. Moreover, it is in the centre of a much less smoky city than London, and most of the members are within a tram ride".

Whichello's suggestion, therefore, was to establish a Loan Fund, with loans being secured from the Association's wealthier members, to enable the purchase of second hand instruments such as 4- to 6-inch (10 to 15 cm) refractors. The borrower would pay a deposit of one-third to one-half of the value of the loaned instrument and pay 5% per year of the remaining value in interest. Whichello considered that the imposition of such an annual fee was "essential to the financial success of the scheme, and would prevent instruments being *kept when not in use*". He also envisaged instruments being donated or bequeathed to the Association for loan. The proposal was reported in the "Scientific News" section of the *English Mechanic*, which noted that it "deserved consideration" (25). This was followed by some lively correspondence in support of the idea in future editions, albeit that the Association seemingly refused to discuss the matter openly. In this respect one disgruntled member alleged "bad faith" on behalf of the Association claiming that "the subject cannot be discussed at the meeting of the Association", for when someone tried to raise the topic at the November 1898 BAA meeting "the President declined to allow it altogether" (26). However, no official response to the suggestion was forthcoming and it was to take several years before a BAA instrument loan collection of was established. Subsequently Whichello himself was to be both a beneficiary of and a donor to the collection, for in the late 1930's and early 1940's he borrowed an equatorial mount (27) and a cylindrical slide rule. He also personally donated a chronometer to the collection (28).

Meanwhile, Whichello equipped himself with a 9-inch (23 cm) Newtonian reflector which he then replaced in 1900 with a 6-inch (15 cm) refractor by Wray on a driven equatorial mount. He built an observatory for the Wray in his back garden at Tattenhall (29), taking great care in aligning the mount. He later wrote up his alignment method in the BAA *Journal* (30). Although the rural setting of his observatory was obviously conducive to astronomy, there was one drawback: "In the country, especially if remote from the rail, it is always difficult to find the right time. In this village the church clock is always wrong. Having no transit instrument, I have often thought of setting up a sundial by which to correct my clocks once or twice a week" (31). Thus he set about constructing a sundial which could be read to one-quarter of a minute. Details and drawings were published in the *English Mechanic* (31). He clearly had a practical inclination as he also designed a modification to his weight-driven equatorial drive to control the drive rate. This "air clock" involved resting the falling weight on an inflated football bladder and controlling its rate of fall by adjusting the release of air from the bladder with a needle valve (32). However, in spite of much experimentation it was never completely satisfactory and was later abandoned. Another project surfaced in 1913 when he was investigating the installation of acetylene lighting at The Mount (33).

After joining the BAA Whichello also began submitting drawings to its Lunar Section; his first observation reported in the *Journal* was of the total lunar eclipse of December 1898 (34). Subsequent drawings appeared in the *Journal*, the BAA *Memoirs* and the *English Mechanic*. Evidently the Lunar Section Director, Walter Goodacre (1856-1938), was delighted with his work noting that "Dr. Whichello has been very active in making sketches of different formations, of which about thirty have been sent in. These illustrate, among other objects, rilles and ridges near Cauchy, the West wall of Hipparchus, Posidonius [Figure 5] and several of Atlas". His line drawing of the Straight Wall is shown in Figure 6.

Whichello remarked: (35) "On Jan. 9, 1900, I was observing the 'Straight Wall'. At the S. extremity is the forked end, which is often compared to a stag's horn handle of a stick. This seemed to me really to be part of a ring crater, the rest of which is ruined. Till tonight [27 April 1900] I have not had a chance to see this again. But at eight o'clock this evening the ruined ring was easily visible".

From as early as 1899, Whichello also submitted regular sunspot drawings to the BAA Solar Section which he continued to do until almost the end of his life. His work featured regularly in Solar Section contributions to the BAA Memoirs. Drawings of large sunspots groups made in 1899, 1903 and 1907 are shown in Figures 7 and 8.

Thus for several years Whichello was most interested in observing the Moon and Sun. However, a new astronomical interest captivated his imagination in 1915 when he began to observe variable stars with his 6-inch Wray, continuing at a steady pace for the next twenty years. The BAA Variable Star Section's database contains over two-thousand of his estimates, the first being of the long-period variable T UMa on 4 January 1915 and the last on 4 August 1935 (36). His most frequently observed star was X Cam, another long-period variable.

In 1900 Whichello joined other BAA members on an expedition to Algiers to observe the total solar eclipse on 28 May. The original plan was for the BAA to charter a ship, much like it had done previously for the Norwegian eclipse of 1896, but the idea was abandoned with the onset of the Boer War in South Africa which resulted in few available ships for touristic charters (37). Several BAA members elected to go to Algiers on the Orient Steamship Company's steam yacht *Argonaut* (38), thus Whichello, Longbottom and another Cheshire amateur astronomer, Harry Krauss Nield (39), decided to join them. The party, comprising some 40 people, disembarked at Cape Mantifou in the Bay of Algiers. The first requirement was to find a suitable observing location. Whichello and Krauss Nield, both French speakers, were despatched to identify a suitable spot for the party and following discussions with the village school teacher, they selected the local school premises.

Memories of the clouded out BAA expedition to Norway in 1896 were uppermost in the minds of most, but fortunately the Algerian eclipse passed off under almost perfect conditions. Whichello was able to sketch the solar corona, noting its extension in the direction of Mercury, some 7.3 lunar radii from the moon's centre, which took the form of an "angel's wing". This feature was recorded by several other observers, including those at other locations, and can been see in Krauss Nield's sketch in Figure 9. By all accounts the experience of totality made a great impression on Whichello and he often referred to it with enthusiasm in later years (40).

## Computational astronomy

In the early 1920s, Whichello began contributing regularly to the BAA Computing Section (41) and from the mid 1930's, as developing health problems interfered with night-time observing, he focused further on computational work. Much of his work was in the prediction and analysis of lunar occultations, which were published in the BAA *Journal* and *Handbook* and in the *Nautical Almanac*. In 1940, with many members away on active service, Dr. J.G. Porter (1900-1981), the Director of the Computing Section noted that "The computing is principally in the hands of K. Pollock and H. Whichello, all the work being in duplicate" (42).

Whichello also supported the Comet Section in calculating ephemerides, usually in association with W.P. Henderson. Particularly noteworthy was their work on Comet 31/P Schwassmann-Wachmann 2. This comet was discovered by A. Schwassmann and A. A. Wachmann at the Hamburg Observatory in Germany on 17 January 1929 and a period of 6.43 years was calculated (43). At its first predicted return in 1935 it was unfavourably placed and although it was eventually recovered by G. Van Biesbroeck, few observations were made during the apparition. Thus the next perihelion passage, in 1942, was eagerly awaited. Whichello and Henderson's ephemeris, based on all the available astrometry from the previous two apparitions, was published in the BAA *Handbook* for 1941 (44). Guided by their ephemeris, Dr. H.M. Jeffers of the Lick Observatory in California recovered the comet on 20 September 1941 at almost the exact spot predicted by Whichello and Henderson. Speaking shortly after this the Reverend Dr. M. Davidson, Director of the Comet Section, told BAA members at the October 1941 meeting that "this is a fine triumph for the computers and reflects great credit on the Computing Section, composed of amateurs, which is capable of such accurate work, and I would congratulate the Section and its work" (45).

Whichello and Henderson also produced ephemerides for the 1942 perihelion passage of 14P/Wolf (46), although their prediction for perihelion was out by 16 days as they did not allow for perturbations by Jupiter and Saturn (47). Their ephemeris for 33P/Daniel's 1943 passage (48) was more accurate and enabled BAA member G.F. Kellaway (1902-1962) to recover the comet visually with a 12-inch (30 cm) reflector at West Coker in Somerset about ½° away from the predicted position (49).

**Local astronomical societies**

In addition being a member of the BAA and a Fellow of the RAS (50), Whichello also maintained links with overseas astronomical societies and in 1923 the Societé Astronomique de France elected him a life member.

F.W. Longbottom had been the driving force in the establishment of an astronomical society in Chester in 1892 under the auspices of the Chester Society for Natural Science, Literature and Art. Whichello joined the Society's Astronomical Section and it was there that he met Longbottom and their friendship began to flourish. Whichello's first of many talks to the Section was in December 1899 when he spoke on "The Spangled Heavens" (51). In 1927 he was presented with the Society's top award, the Kingsley Memorial Medal (52), for his work on variable stars.

Whichello was also active in the Liverpool Astronomical Society for many years. He served as Honorary Secretary between 1914 and 1922, going on to be President between 1923 and 1925 and then again between 1931 and 1935. He also served on the committee of Manchester Astronomical Society and he spoke many times at meetings of both societies (53).

**Later years**

Around the time of his move from Heswall to Chester in 1923, Whichello took early retirement from medicine as a result of a hearing impediment (54). His wife died in 1934 and in his later years he lived with his younger daughter, by then widowed, on Liverpool Road near the centre of Chester. Finding it difficult to use his telescope, due to his increasing physical limitations, he put the 6-inch Wray up for sale in 1936, (55) thus marking the end of his night-time observations. As we have seen, from then on he concentrated on the computational side of astronomy, although he did later obtain a 3-inch (7.5 cm) refractor which he used for solar work.

During World War 2, Whichello prepared summaries of papers from overseas astronomical journals, such as *Popular Astronomy* and *Proceedings of the National Academy of Sciences*, which were published in the BAA *Journal*. These were greatly appreciated by members as such publications were not widely available in Britain due to difficulty of sending them across the Atlantic from the USA. Moreover, with many members away on active service, the submission of other material to the Journal was significantly reduced (56).

On 16 May 1945, only 8 days after VE celebrations, Whichello passed away at his home, aged 75. His funeral took place at Chester Cathedral two days later. Obituaries were published in the BAA *Journal* (40), the *Chester Courant* (57) and the *Chester Observer* (58). The BAA obituary was written by longstanding BAA member Dr. R.W. Eldridge who as a schoolboy in 1917 had contacted Whichello in his capacity as Secretary of the Liverpool Astronomical Society (59). Always keen to encourage others, Whichello invited Eldridge to his home at Heswall and allowed him to use his telescope on many occasions. Eldridge noted (40) that "one of his most delightful characteristics [was] his desire to fan the flame of any youthful love of science which he came to hear about, and to stimulate this interest in every possible way. Inquiries from strangers were nearly always followed by practical help, an invitation to his home, an evening with his telescope, an astronomical discussion and an introduction to his scientific friends".

Other obituaries commented on Whichello's engaging personality adding that "Dr Whichello was a charming man and a fluent conversationalist" (57) and "he was noted for his charm, his courtesy, his innumerable anecdotes and his generosity. No one was more modest about his learning" (58). It was also mentioned that he was fond of sketching and that he was a fresh air and swimming enthusiast, even to the extent of swimming regularly in the River Dee at Chester during the winter months.

## Reflections

The success and international reputation of the BAA over its 120 year existence can largely be attributed the varied contributions of its members to its observing sections. Whilst the names of many pioneering members are still cherished, others have slipped from memory. Dr. Harold Whichello is one such name. The Association attracted individuals from all socio-economic backgrounds, irrespective of gender, but its membership in the early years was particularly well represented by clergymen, military officers and medical practitioners, such as Whichello. His professional activities provided the financial security which enabled him to purchase a relatively large telescope as well as assuring him sufficient leisure time to pursue a hobby which, for him, became a life-long interest and during which he made important contributions to the work of the Association.

## Acknowledgements

The authors are most grateful for the assistance we have received from a great many people whilst preparing this paper. We acknowledge Gerard Gilligan and the Liverpool Astronomical Society for giving permission to publish the photograph of Whichello. Richard Wildman, Bedford Modern School Archivist, provided information about Harold and Norman Whichello's attendance at the school. Bill Leatherbarrow made helpful comments on Whichello's lunar observations and Bob Marriott on the instruments loaned from and donated to the BAA. Roger Lane provided information about Whichello's medical activities in Heswall and Jennifer Higham, RAS Librarian, gave details of Whichello's RAS Fellowship.

This research made extensive use of scanned back numbers of the *Journal*, which exist largely thanks to the efforts of Sheridan Williams; similarly we used the scanned archived of the *English Mechanic*, supplied by Eric Hutton. The BAA Variable Star Section's online database, the NASA/Smithsonian Astrophysics Data System, the British Newspaper Archive (British Library) and the Cheshire County Archives were also accessed in the preparation of this paper.

Finally we thank our referees for their constructive comments.

## Addresses

JS: "Pemberton", School Lane, Bunbury, Tarporley, Cheshire, CW6 9NR [bunburyobservatory@hotmail.com]

TH: "The Hawthorns", Burwardsley Road, Tattenhall, Cheshire, CW3 9NS

2. Stroobant P., Delvosal J., Philippot H., Delporte E. & Merlin E., Les Observatoires Astronomiques et les Astronomes, publ. Royal Observatory of Belgium (1907); compilation of the contents began in 1902.

3. UK contacts included R. Killipp and R.G. Johnson of Liverpool Astronomical Society, E. T. Whitelow of Manchester Astronomical Society, T.H.E.C. Espin of the Newcastle-upon-Tyne Astronomical Society, A. Mee of the Astronomical Society of Wales. The Directors of the following BAA Sections were also contacted: Solar (A.L. Cortie), Lunar (W. Goodacre), Jupiter (T.E.R. Phillips), Saturn (G.M. Seabrooke), Comet (E.W. Maunder was contacted although the Director was actually A.C.D. Crommelin), Variable Star (E.E. Markwick) and the West of Scotland Branch (A.D. Ross was contacted; he would be a later Branch President; the President at the time was actually Robert Robinson). Whether other Sections or Branches were contacted or whether no response was received is not known. Overseas societies were also contacted, such as the Societé Astronomique de France, the Astronomical and Astrophysical Society of America and the Royal Astronomical Society of Canada.

4. Amy Whichello wrote a history of Wollaston. The third edition published posthumously in 1930 has a Preface by Harold Whichello. "Annals of Wollaston", publ. Albert W. Green, Wollaston (1930).

5. Norman Whichello later studied electrical engineering. In the 1891 Census both Harold and Norman were recorded as living at 3 Hill View, Croydon.

6. Bedford Modern School archives. Whichello was a member of Langley House, a boarding house. His brother Norman also attended the school between 1882 and 1888. Norman was apparently viewed as the more academic of the two, and went on to achieve a First in the City & Guilds of London Institute (Engineering Dept) in 1890.

7. Whichello H., The Lancet, 28 July 1894, page 224.

8. Creosote, more commonly used in recent times to treat wood, but now outlawed in many countries due to its toxicity, was a popular treatment for TB at the time. Whichello noted that the creosote was administered "per rectum", which could not have been a pleasant experience either for the patient or the person administering it.

9. The Mount was built in 1842.

10. The Read family lived at 'Kenwyn', Dee Banks, a well-to-do part of Chester, and the wedding took place in St. Paul's Church, Great Boughton, Chester. Blanche Mary Read was born at Fairfield, Liverpool.

11. Aged 7 years in the 1911 Census. She married Humphrey Gaskell in 1930 and lived in Chester.

12. Aged 5 years in the 1911 Census. She married William Blake-Dyke in 1927 and lived in Chester. In later years, after the death of his wife, Whichello lived with Eva Florence.

13. Whichello's parents lived at "Sunnyside", just along the High Street from "The Mount", on the opposite side of the road. They spent the rest of their lives in Tattenhall and were buried in the churchyard of St. Alban's Church. Upon their father's death in 1908, each son received a legacy of £1000.

14. The Chester Observer, 20 May 1899.

15. Whichello wrote a summary of the incident for *The Lancet*, which contains a more detailed medical analysis. He noted that George's watch was still working. Whichello H., The Lancet, 3 June 1899, page 1490.

16. Whichello's description of the case and his objections to the use of the metric system in prescribing medicines are described in: Whichello H., British Medical Journal, 9 April 1904, page 366.

17. He named his house on Thurstaston Road, Heswall, "Linslade" after his birth place. A house of that name no longer exists. Heswall was in the County of Cheshire until the infamous 1974 boundary changes. It is now in Merseyside

18. The first edition was published in 1885.

19. Whichello was elected at the BAA meeting of 26 January 1898 (JBAA, 8, 199 (1898)). His candidature was announced in JBAA, 8, 141 (1898).

20. A biography of Longbottom has been prepared: Shears J., JBAA, accepted for publication (2012).

21. Schooling was BAA Secretary from 1892 to 1893 and again from 1896 to 1900.

22. Whichello H., JBAA, 9, 76-77 (1898).

23. The proposed BAA observatory and the origins of the Association's instrument collection is described in Marriott R.A., JBAA, 117, 309 (2007).

24. Whichello H., English Mechanic, 1764, 517 (1899).

25. English Mechanic, 1762, 467 (1898).

26. This statement was made in an anonymous letter to the English Mechanic by "A Member of the B.A.A."; English Mechanic, 1765, 540 (1899).

27. BAA Instrument number 38; "Brass equatorial head, suitable for 4-inch refractor". Loaned to Whichello in 1938 (JBAA, 48, 400 (1938)).

28. BAA Instrument number 70: 8-day chronometer by James Hatton of London. Presented to the Association in 1938 (JBAA, 48, 299 (1938)). The instrument was promptly loaned to B.M. Peek (JBAA, 48, 400 (1938)). Bob Marriott, Director of the BAA Instruments & Imaging Section, notes that both the chronometer and the equatorial mount were sold by the BAA by auction at Sotheby's in 1987.

29. Apparently the 9-inch reflector was only used for a short time at Tattenhall. When Whichello later moved to Heswall he built a new 12 ft dome. In a 1917 letter (Whichello H., English Mechanic, 2713, 168 (1917)) he mentioned that he had been using the new dome for 3 years, i.e. since 1914.

collection for several years was loaned to Eldridge. As mentioned elsewhere, the Association sold the mount in 1987.

OBSERVATOIRE ROYAL DE BELGIQUE

SERVICE ASTRONOMIQUE

LES

# OBSERVATOIRES ASTRONOMIQUES

ET LES

## ASTRONOMES

PAR

P. STROOBANT
ASTRONOME

J. DELVOSAL et H. PHILIPPOT
ASTRONOMES ADJOINTS

E. DELPORTE et E. MERLIN
ASSISTANTS

du Service astronomique de l'Observatoire royal de Belgique.

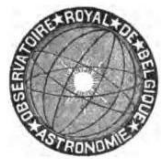

BRUXELLES

HAYEZ, IMPRIMEUR DE L'OBSERVATOIRE ROYAL DE BELGIQUE

Rue de Louvain, 112

1907

(a)

**Chester** (Angleterre).

Longbottom, F. W., Haslemere, Queen's Park, Chester.

Photographie de comètes et de nébuleuses.

Réflecteur équatorial de Cope, miroir de 306 millimètres et 0ᵐ61 de foyer.

Whichello, H., Dʳ. The Mount, Tatterhall, Chester.

· Lune.

Réfracteur de 152 millimètres d'ouverture.

(b)

Figure 1: "*Les observatoires astronomiques et les astronomes*" (a) title page (b) entry for Chester
(note: *Tattenhall* is misspelt)

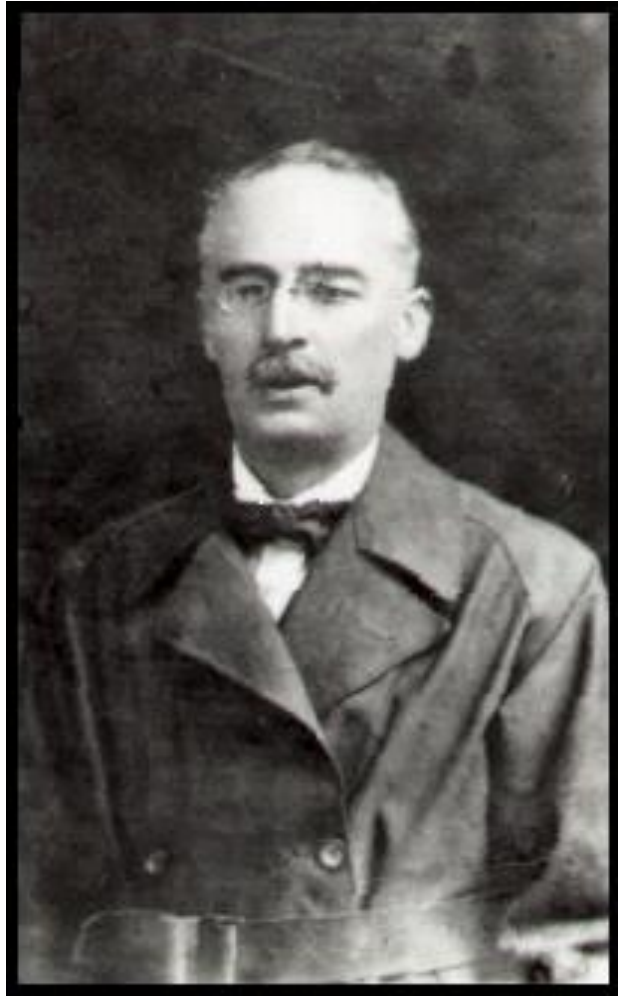

Figure 2: Dr. Harold Whichello (1870-1945)

(image courtesy of the Liverpool Astronomical Society)

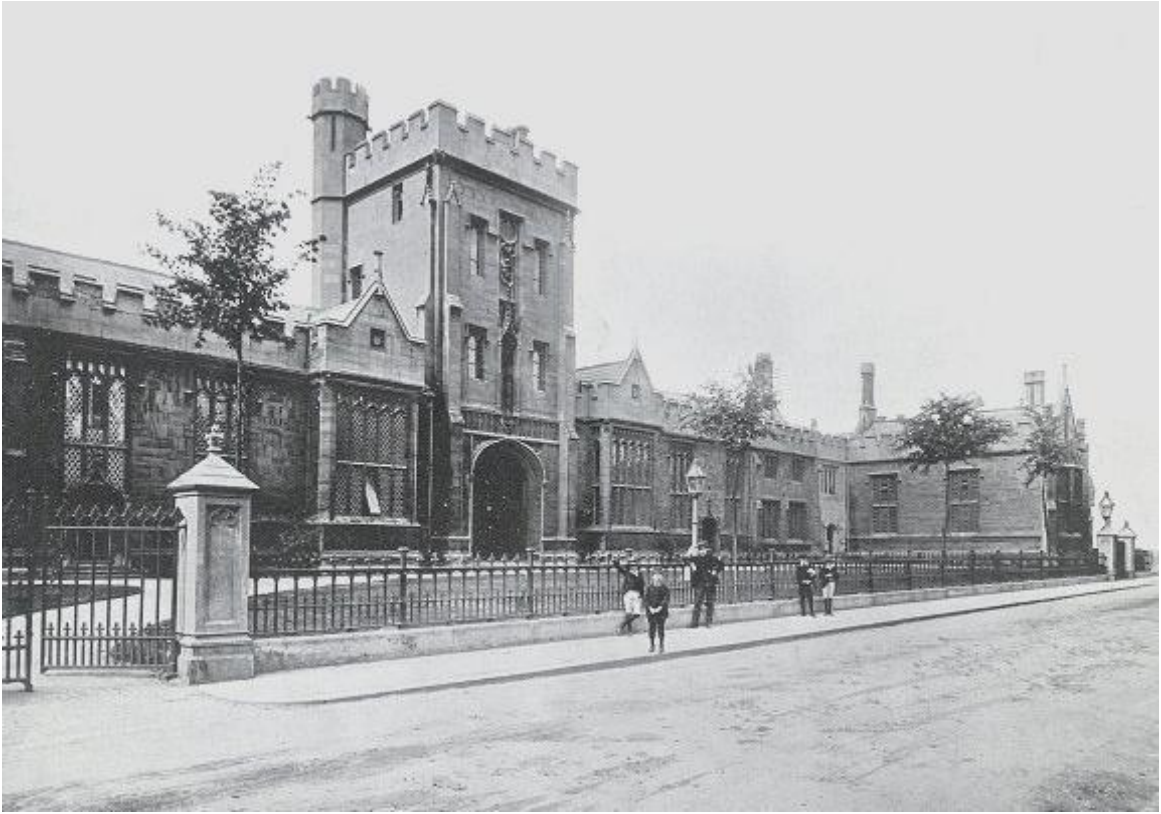

Figure 3: Bedford Modern School

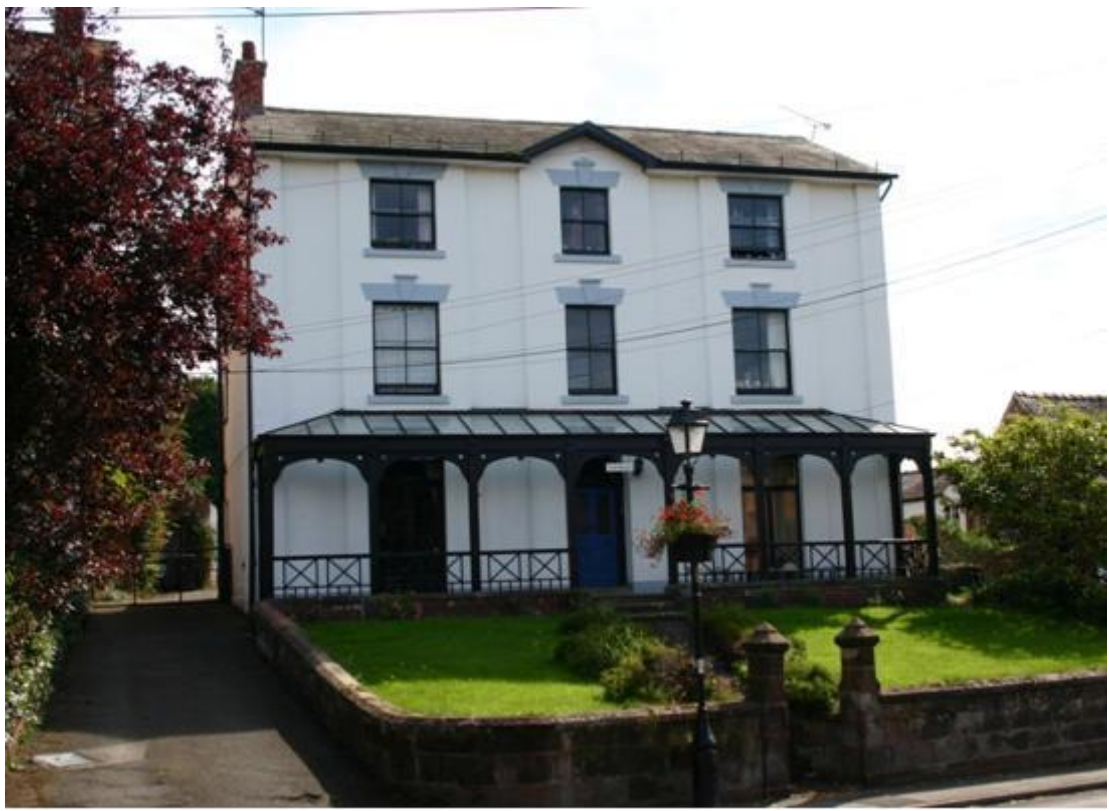

Figure 4: "The Mount", Whichello's residence in Tattenhall (Jeremy Shears)

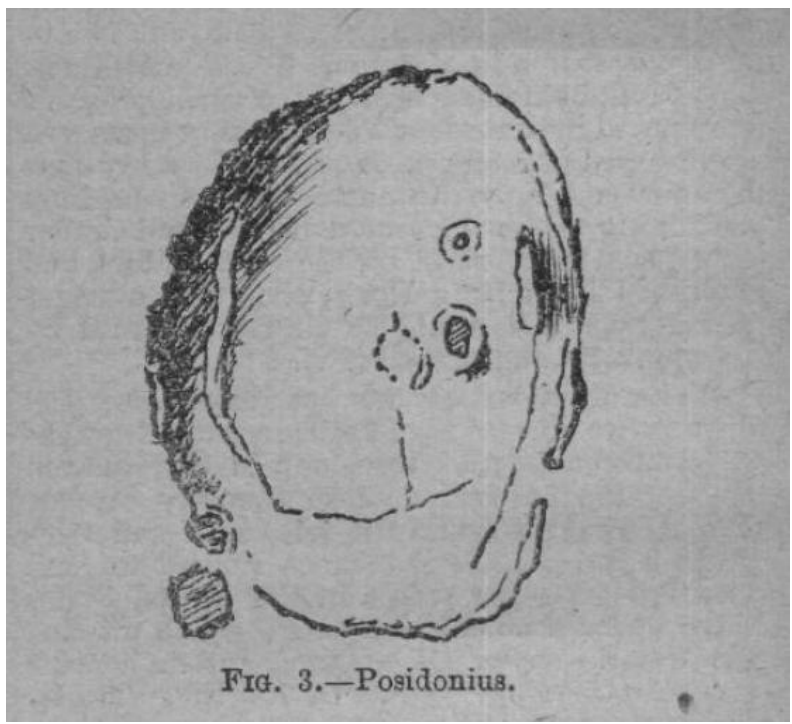

Figure 5: The lunar crater Posidonius (1903). From reference  (60)

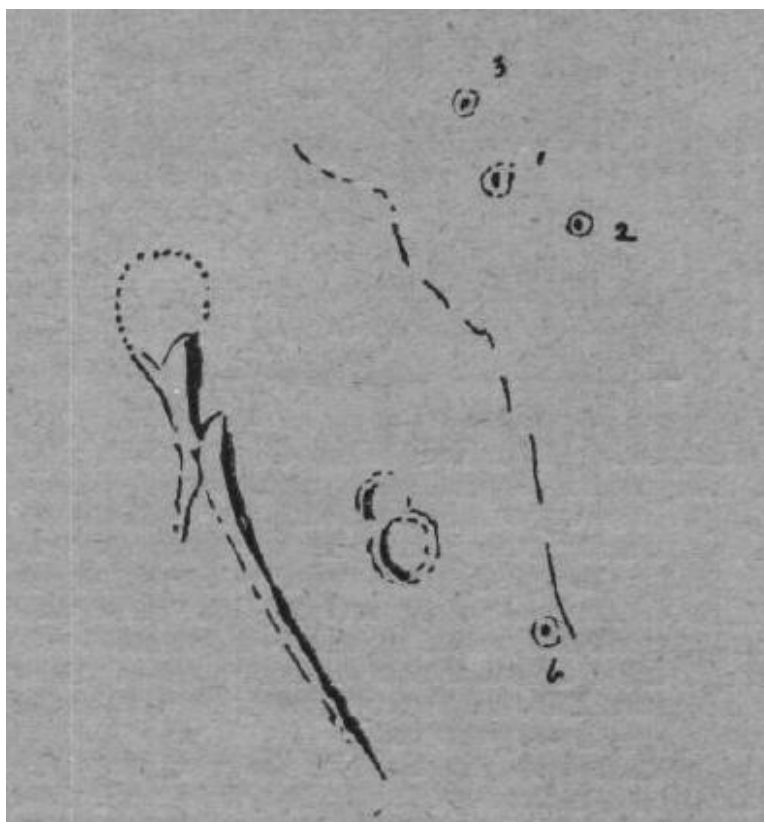

Figure 6: The Straight Wall.

The dotted circle at the end of the stag's horn feature is a partial "ghost ring" which
Whichello suspected was a flooded crater in the *Mare Nubium*. 27 April 1900, 6 inch
refractor. From reference (35)

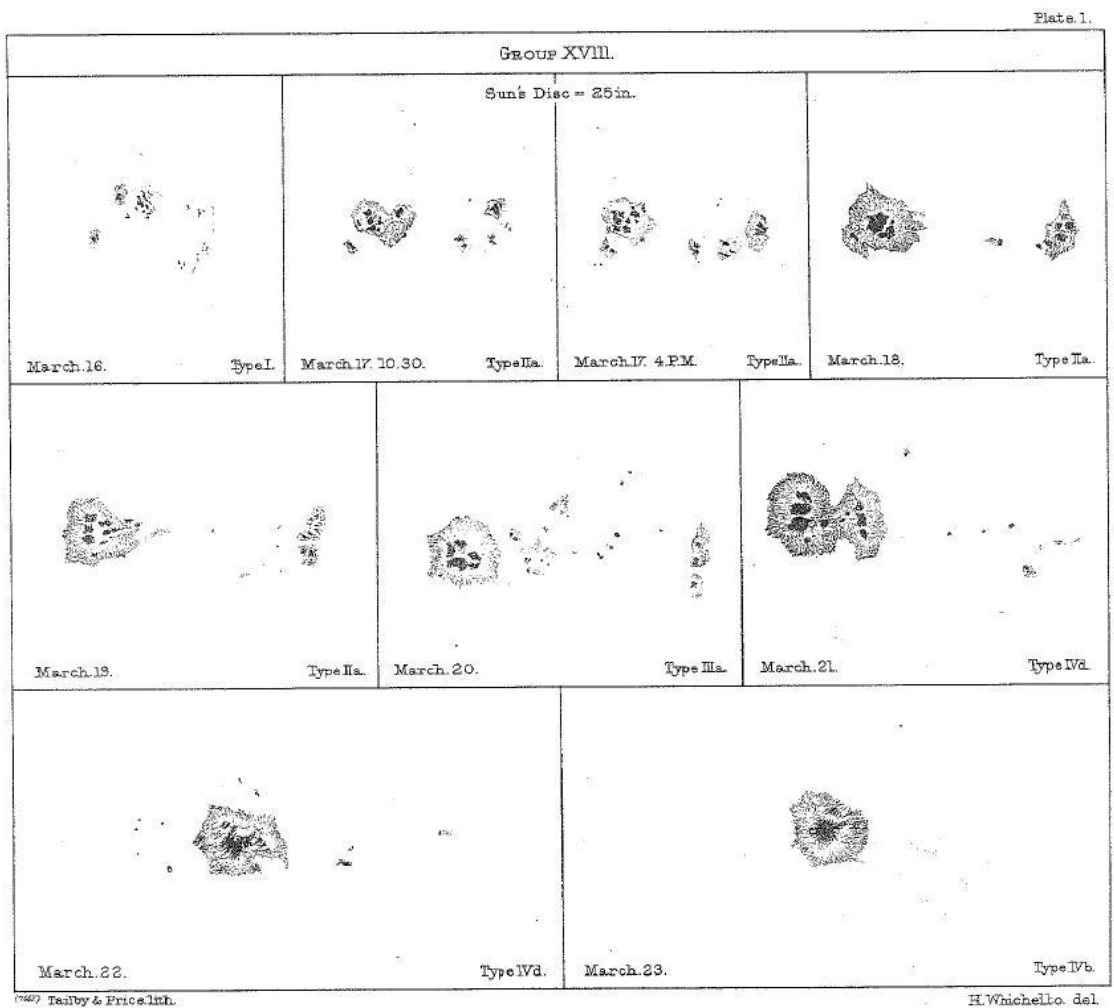

Figure 7: Evolution of a sunspot group during March 1899 (61)

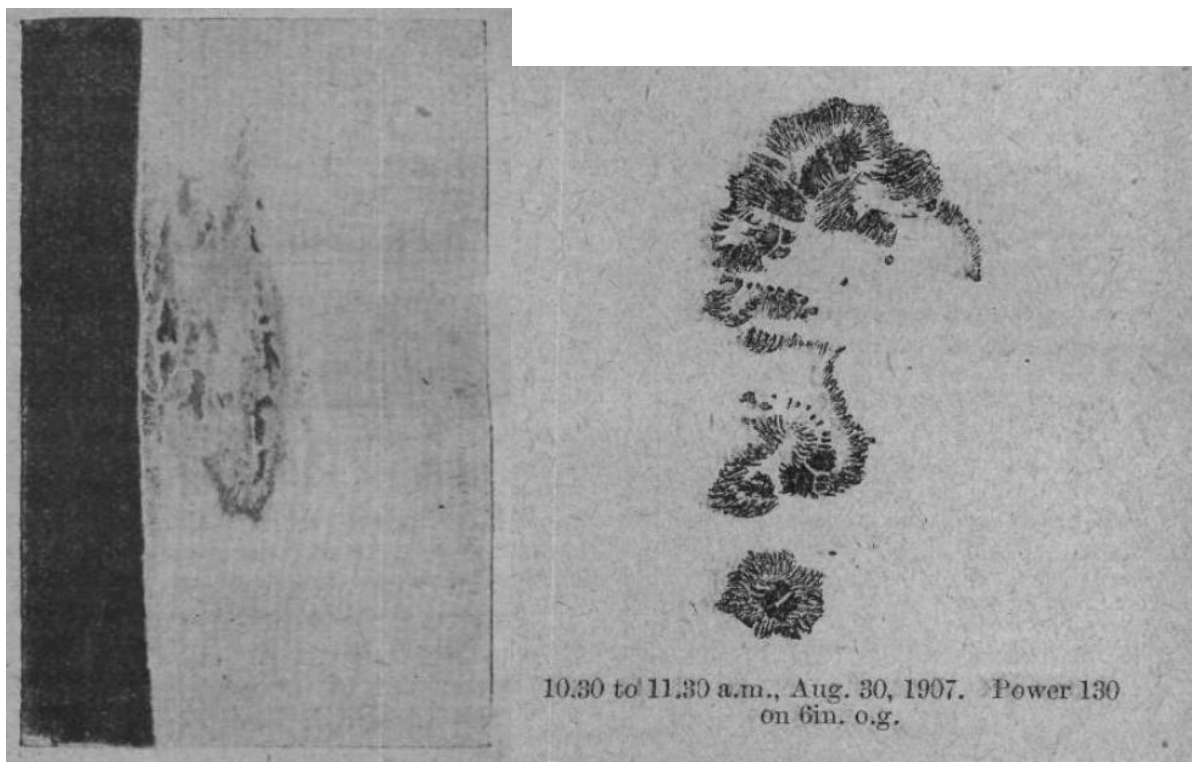

Figure 8: Sunspots drawn using the 6-inch Wray refractor

Left - 17 October 1903. From reference (60)

Right- 30 August 1907. From reference (62)

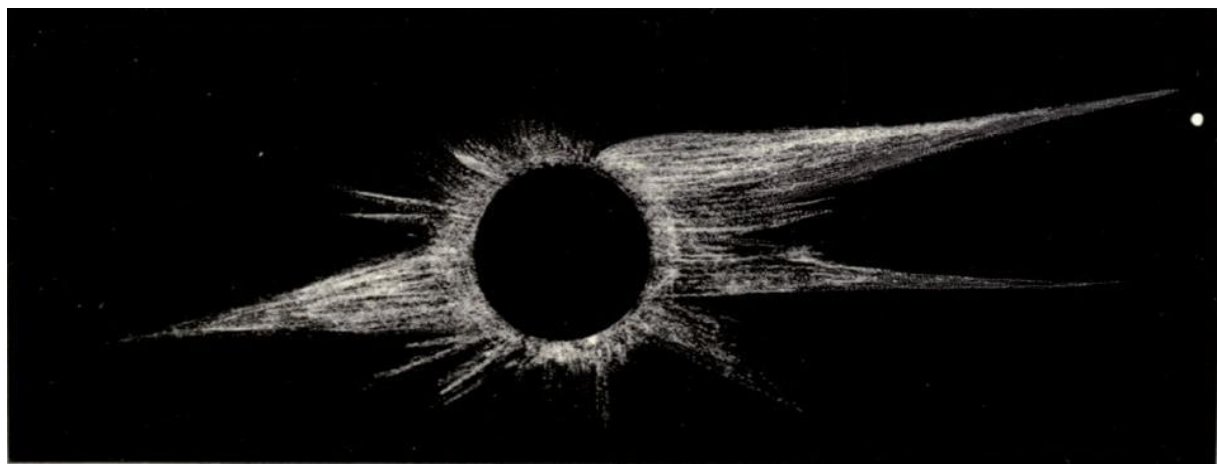

Figure 9: The eclipse of 28 May 1900 drawn by H. Krauss Nield at Algiers, showing the distinctive "angel's wing" form of the corona projecting towards Mercury. Note that this is a combined sketch made from sketches contributed by several observers at the Cape Mantifou site. From reference (37)